\documentclass[12pt]{article}
\usepackage{graphics}
\begin{document}
\begin{center}
{\large Calculation of Spin Observables for Proton-Neutron Elastic Scattering in the Bethe-Salpeter Equation}
\end{center}
\begin{center}
{Susumu Kinpara}
\end{center}
\begin{center}
{\it National Institute of Radiological Sciences \\ Chiba 263-8555, Japan}
\end{center}
\begin{abstract}
Bethe-Salpeter equation is applied to $p$-$n$ elastic scattering.
The spin observables are calculated by the M matrix similar to $p$-$p$ case.
The parameters of the meson-exchange model are used with the cut-off for the pion exchange interaction.
Change of the M matrix indicates breaking of the charge independence in the nucleon-nucleon system.
\end{abstract}
\hspace*{4.mm}
Recently proton-proton ($p$-$p$) elastic scattering has been treated 
in the framework of the Bethe-Salpeter (BS) equation$\cite{Kinpara}$.
Expanding the BS amplitude by a set of the $\Gamma$ matrix the resulting simultaneous equations 
to solve the expansion coefficients describe  
not only the bound state (deuteron), also the scattering state of nucleon-nucleon two-body system. 
\\\hspace*{4.mm}
One of the characteristics of the present formulation is the breaking of the charge independence
expressed by the disagreement of the isospin $T$=1 part of the M matrix between $p$-$p$
and $p$-$n$ system ($M^1_{p-p} \neq M^1_{p-n}$). 
Here, the electro-magnetic interaction is turned off from the outset 
and then the inequality works independent of the Coulomb force.
The origin of the $T_z$ dependence is attributed to the form of the interaction potential 
in which the pseudovector coupling constant of pion is required to change from the standard value $f \sim $ 1 to the 
$\sim 1/{\rm 20}$ to reproduce the experimental data on $p$-$p$ elastic scattering ($T_z$=1).
In addition to it, the higher-order correction of the ${}^3 P_2$ state is essential and therefore necessarily introduce the three-body intermediate state $\pi^++p+n$ to obtain the meaningful value of the phase shift parameter.
\\\hspace*{4.mm}
The many-body process may be described by the higher-order diagrams for the irreducible kernel including  
the lowest-order ladder approximation to give the possible potential terms in the BS equation.
Consequently the virtual $p+n$ state is allowed to contruct the ${}^1 P_1$ scattering wave 
in conjunction with one $P$-wave pion. 
So the $T_z$=0 state plays a decisive role in the two nucleon system irrespective of the initial value of $T_z$.
It is our purpose of the present study to elucidate the elastic scattering of $p$-$n$ system 
at the intermediate energy region. 
\\\hspace*{4.mm}
To obtain the M matrix which determines the dynamical properties between two nucleons
and moreover investigate the possible nuclear force 
the spin observables are indispensable following the differential cross section.
Although the number of the experimental data is not as many as that of the $p$-$p$ elastic scattering,
calculation of the $p$-$n$ spin observables in terms of the M matrix formulation is still feasible and makes us 
draw some conclusions on the two-body interaction from the results.
\\\hspace*{4.mm}
While in the lowest-order ladder process the $p$-$p$ elastic scattering does not employ exchange of the charged pions 
to generate the nuclear interaction 
and the pseudovector coupling constant of pion is reduced from the standard
value, the $p$-$n$ case is seen that the values of the set of the meson exchange parameters follow 
approximately the same ones used for the finite nuclei and the dependence on the incident energy is much weaker
particularly at the low energy region.
Then the ladder approximation for the irreducible kernel is appreciated and also the three-body state is supposed
to correct it for the investigation of the elastic scattering in detail.
\\\hspace*{4.mm}
Expanding the BS equation by the $\Gamma$ matrix and imposing the auxiliary condition the resulting simultaneous 
equations are the form equivalent to the nonrelativistic Schr$\ddot{\rm o}$dinger equation 
under the nucleon-nucleon potential.
In order to evaluate the higher-order correction beyond the Born approximation the inverse square part of the potential
is remained and the solution of the scattering state is substituted to evaluate the element of the two-body interaction 
relating to the initial ${}^3 S_1$ state.
Each element of the interaction potential of pion is shown as \\
\begin{equation}
<L^\prime \mid V(\mit\Lambda_{\rm 0}) \mid \nu>\;=\;<L^\prime \mid V(\mit\Lambda_{\rm 0}) \mid L> F(\mit\Lambda),
\end{equation}
\begin{equation}
F(\mit\Lambda) \equiv F_{\rm 0}(\mit\Lambda) \cdot F_{\rm 1}(\mit\Lambda),
\end{equation}
\begin{equation}
F_0(\mit\Lambda) \equiv \; <L^\prime \mid V(\mit\Lambda) \mid \nu_{\rm 0}>/<L^\prime \mid V(\mit\Lambda) \mid L>,
\end{equation}
\begin{equation}
F_1(\mit\Lambda) \equiv \; <L^\prime \mid V(\mit\Lambda) \mid \nu>/<L^\prime \mid V(\mit\Lambda) \mid \nu_{\rm 0}>.
\end{equation}
Here, $\mid \nu >$ and $\mid \nu_{\rm 0} >$ denote the exact solution 
and the approximate one under the inverse square potential respectively.  
The value of the cut-off parameter $\mit\Lambda_{\rm 0}$ for the Born term 
is fixed at $\mit\Lambda_{\rm 0}\sim {\rm 500}$ MeV.
It is verified that $F_0(\mit\Lambda)$ is finite as the cut-off $\mit\Lambda \rightarrow \infty$ and which 
we actually use in the present calculation.
We do not attempt to draw some conclusions on the value of $F_1(\mit\Lambda)$ here
and assume $F_1(\mit\Lambda)\approx{\rm 1}$ for the sake of simplicity.
The higher-order correction of the ${}^3 S_1$ state entails the change of the isospin $T$=$0$ parts $M^0_{11}$ and $M^0_{00}$ 
of the the M matrix through the phase shift $\delta_{01}$ and the mixing parameter $\epsilon_1$ of the $(L,J)=(0,1)$ state
influenced by $F_0(\mit\Lambda)$ and $F_1(\mit\Lambda)$.
\\\hspace*{4.mm}
One of the interesting subjects in the BS formalism for elastic scattering is the spin singlet ($S$=0) wave
for which both of the axial-vector ($av$) and the pseudoscalar ($ps$) equations are applicable.
In our previous study$\cite{Kinpara}$ the spin observables are calculated for $p$-$p$ case 
by using each equation individually and they have been compared with the experimental data.
For $p$-$n$ case the numerical results with the $av$ and $ps$ equations 
are shown in the present study making us choose one $S$=$0$ component among them.
\\\hspace*{4.mm}
The formulation of the spin observable is performed by using the density matrix in which the quantity
$a_{\mu\nu\tau\rho}^T$ is defined as \\
\begin{equation} 
a_{\mu\nu\tau\rho}^T \equiv 
\frac{1}{4}\, {\rm Tr}\, M^T \sigma_\tau^{(1)}\sigma_\rho^{(2)} {M^T}^\dagger \sigma_\mu^{(1)}\sigma_\nu^{(2)}
\;\;\;\;\;\;(T=0,1)
\end{equation}
where $\sigma_\mu^{(i)}$ ($i=1,2$) is same as in ref. [1].
The description of the $p$-$n$ elastic scattering requires the isospin $T$=0 component of the M matrix $M^0$ 
in addition to $M^1$ by changing the sum from $\sum_{T=1}$ in the $p$-$p$ case to $\frac{1}{2}\sum_{T=0,1}$ 
to calculate each spin observable.
The labels $\mu$, $\nu$, $\tau$ and $\rho$ in Eq. (5) denote the directions of spin for
the scattered, the recoil, the incident and the target nucleons accordingly.
\\\hspace*{4.mm}
For $p$-$n$ elastic scattering the ladder approximation works well as illustrated 
by comparing results of the calculation with and without the higher-order correction to the Born term.
In it the approximate wave function by the leading inverse square potential part is substituted for the exact one.
Like the bound state the ${}^3S_1$ partial wave is expected to play a significant role 
in the scattering state and so we use the procedure for calculating the ${}^3S_1$ state
and also for the ${}^3P_2$ state tentatively to improve the numerical results of the spin transfer parameters.
\\\hspace*{4.mm}
The geometry of scattering is as follows.
The elastic scattering in the center of mass system is specified 
by the momentum of the incident nucleon $\vec k$ 
(parallel to $\hat{z}$ direction of the $(\hat{x}, \hat{y}, \hat{z})$ right-handed system),
the scattered nucleon $\vec k^\prime$ ($\vert\vec k^\prime\vert=\vert\vec k\vert$) and the scattering angle $\theta$.
The unit vector $\hat{n}\, (=\hat{y})$ is defined 
by $\hat{n}\equiv\vec{k}\times\vec{k^\prime}/\vert\vec{k}\times\vec{k^\prime}\vert$.
For evaluating the spin rotation and the spin transfer parameters the unit vectors $\hat{K}$ and $\hat{P}$ are defined by
$\hat{K}\equiv(\vec{k^\prime}-\vec{k})/\vert\vec{k^\prime}-\vec{k}\vert$
and $\hat{P}\equiv(\vec{k^\prime}+\vec{k})/\vert\vec{k^\prime}+\vec{k}\vert$.
They construct the right-handed system $(\hat{K},\hat{n},\hat{P})$.
\\\hspace*{4.mm}
Result of the differential cross section $I_0(\theta) \equiv \frac{1}{2}\sum_{T=0,1} a_{0000}^T$ is shown 
as a function of the center of mass scattering angle $\theta$ at the 200 MeV laboratory energy in Fig. 1.
Similar to $p$-$p$ elastic scattering
the size of the result of the $ps$ equation becomes larger than that of the $av$ equation owing to the difference
of the interaction for the $S$=0 wave.
Thus inclusion of the isospin $T$=0 part of the M matrix maintains the characteristics of the theoretical curves.
In the case of the $av$ equation the deviation from the experimental data$\cite{Keeler}$ has $\theta$ dependence 
apparent particularly at ${\rm cos}\,\theta=\pm {\rm 1}$ suggesting the higher-order correction of the $P$-wave
as discussed later on.
\\\hspace*{4.mm}
The polarization $P(\theta)$ and the analyzing power $A_y(\theta)$ are equivalent and which are
given by $(P(\theta),\,A_y(\theta))\equiv\frac{1}{2}\sum_{T=0,1} (a_{n000}^T,\,a_{00n0}^T)/I_0$ in the present formulation.
As well as the $p$-$p$ case the symmetric relation of the polarization $P(\pi-\theta)=-P(\theta)$ exists 
due to the exclusion principle for identical two fermions.
Fig. 2 shows the results of the calculation of $P(\theta)$ for the $av$ and $ps$ equations 
at the 200 MeV laboratory energy as a function of the center of mass scattering angle $\theta$.
Generally these curves describe the trend of the measurement$\cite{Clough}$.
The remaining difference may be reduced by the $P$-wave correction as well as that of $I_0(\theta)$.
Concerning the magnitude of $P(\theta)$ the $av$ is larger than $ps$ 
since $P(\theta)$ is in inverse proportion to $I_0(\theta)$.
\\\hspace*{4.mm}
Fig. 3 shows the result of calculations for the spin correlation parameter
$A_{yy}(\theta) \equiv \frac{1}{2}\sum_{T=0,1} a_{00yy}^T/I_0$ as a function 
of the center of mass scattering angle $\theta$ at the 200 MeV laboratory energy.
It is obviously seen that the result is different between the $av$ and $ps$ equations.
The $ps$ gives the values of $A_{yy}(\theta)$ much lower than 
that of the $av$ case and the experimental data$\cite{Bandyopadhyay}$
since in the $ps$ equation the interaction potential of the $(S,T)$=(0,1) part 
gives rise to the contribution of $M^1_{ss}(\theta)$ strongly at $\theta=90^\circ$.  
As seen in our previous study$\cite{Kinpara}$ the situation is similar to the $p$-$p$ elastic scattering although
the value of the pion-nucleon pseudovector coupling constant $f$ adopted is about $1/20$ 
of the standard value $f \sim 1$ used for the calculation of the $p$-$n$ elastic scattering.  
\\\hspace*{4.mm}
It has not been answered that the dependence of $A_{yy}(\theta)$ at $\theta=90^\circ$ on the incident energy of nucleon 
is too strong to keep it in the positive value contrary to the experimental fact at the intermediate energy.
Fig. 4 shows that both curves decrease drastically until they approach to $\sim -1$.
The $T$=$0$ component of $I_0(1-A_{yy}(\theta))$ at $\theta=90^\circ$ becomes to be $\sim |M^0_{11}+M^0_{1-1}|^2$ 
and hence the disagreement indicates that the treatment of the $S$-wave 
by means of the inverse square potential approximation does not attain to explain 
the trend of the energy dependence entirely.
\\\hspace*{4.mm}
In order to reproduce the experimental data $F(\mit\Lambda)$ may be modified from the 
original value by introducing a factor $c$ as $F(\mit\Lambda) \rightarrow c \, F(\mit\Lambda)$.
For the $av$ equation the suitable value of $c$ is about 1.2 at the 500 MeV laboratory energy 
shifting the value of $A_{yy}(\theta=90^\circ)$ to $\sim 0.2$ 
close to the experimental data $\sim 0.1$ at the 425 MeV laboratory energy$\cite{Bandyopadhyay}$.
Here it is noted that the value of $c$ required is fluctuated easily by the fitting procedure for $F(\mit\Lambda)$.
The enhancement of $F(\mit\Lambda)$ by $c$ is attributed to $F_0(\mit\Lambda)$ or $F_1(\mit\Lambda)$.
For example, the component of the Bessel function part could be increased by allowing the Neumann function 
to enter in the $S$-wave under the inverse square potential to some extent.
Besides the correction of $F_1(\mit\Lambda)$ may be done by the perturbative treatment of the residual interaction arising from the mass of pion in the $S$-wave scattering state.
\\\hspace*{4.mm}
Calculation of the other spin correlation parameters $A_{zz}(\theta)$ and $A_{xx}(\theta)$ is also useful 
to examine whether the $S$-wave in the $T$=$0$ part corrected is appropriate or not 
in addition to the case of $A_{yy}(\theta)$. 
The spin correlation parameter $A_{zz}(\theta) \equiv \frac{1}{2}\sum_{T=0,1} a_{00zz}^T/I_0$ is calculated with $c=1$
for the $av$ and $ps$ equations at the 500 MeV laboratory energy in Fig. 5. 
The $av(\pi/2)$ and $ps(\pi/2)$ curves take into account the resonance effect which modifies the spin singlet $P$-wave
as $\delta_1 \rightarrow \delta_1 \pm \pi/2$.
In spite of $c=1$ at $\theta=90^\circ$ the result of the $av$ equation works well and which indicates 
the ${}^3 S_1$ state in $A_{zz}(\theta)$ is not crucial unlike the case of $A_{yy}(\theta)$.
Using the effective value $c=1.2$, $A_{zz}(90^\circ)\sim 0.2$ a rather smaller than the experimental value$\cite{Ditzler}$.
The dependence on $c$ is much slower than that of $A_{yy}(90^\circ)$ because in $A_{zz}(\theta)$ the role of 
the $M^0_{11}$ component is not central as verified from the form $I_0(1-A_{zz}(\theta))$ which is independent of $M^0_{11}$.
\\\hspace*{4.mm}
The spin correlation $A_{xx}(\theta)$ is defined by $A_{xx}(\theta) \equiv \frac{1}{2}\sum_{T=0,1} a_{00xx}^T/I_0$ 
and the results of calculations are shown in Fig. 6.
The existing peak at $\theta=90^\circ$ and the periodic behavior is ascribed to the term $ \sim {\rm cos}\, 2\theta $ 
stemmed from $P_2^2(\theta)$ in $M^0_{1-1}$ as well as $A_{yy}(\theta)$.
In the case of $av$ equation the rising of the curve at $\theta > 130^\circ$ is explained by the $L$=$2$ component  
of $M_{ss}^1(\theta)$ giving the term $\sim {\rm cos}\, 4\theta$. 
The experimental data is seen to be around $A_{xx}(\theta) \sim 0.1$$\cite{Shima}$ extensively 
and by using the effective value $c=1.2$ the result of $av$ is improved as $A_{xx}(90^\circ) \sim 0.2$.
\\\hspace*{4.mm}
As seen in Fig. 7 results of the depolarization $D_{nn}(\theta) \equiv \frac{1}{2}\sum_{T=0,1} a_{n0n0}^T/I_0$ 
for the $av$ and $ps$ equations are different from each other and it is related to
the spin isospin $(S,T)$=(0,0) component of the force which is essential 
to interpret the experimental data$\cite{Warner,Barlett}$.
Then the trend of the measured results such as the sufficient magnitude at $\theta \le 90^\circ$ and the sudden decrease
at $\theta > 90^\circ$ is explicable to a degree by the angular dependence of the $P$-wave ($P_1(\theta)$) 
in $M^0_{ss}(\theta)$.
On the other hand the $S$-wave in $M^1_{ss}(\theta)$ is less important since
the two theoretical curves cross each other at $\theta \sim 90^\circ$.
In the $av$ equation the lack of the magnitude is corrected by the factor $c=1.2$, for example, 
the maximum value at $\theta \sim 75^\circ$ changes to $\sim 0.8$ 
perhaps owing to the cross term $\sim {\rm Re}\,[M^0_{ss}(\theta) \, M^0_{11}(\theta)^\ast$] in $I_0 D_{nn}(\theta)$.
\\\hspace*{4.mm}
The quantities $(K_{ij}(\theta),D_{ij}(\theta)) \equiv \frac{1}{2}\sum_{T=0,1} (a_{0ji0}^T,a_{j0i0}^T)/I_0\;(i,j=x,y,z)$
are mutually exchangeable by $M_{ss}(\theta) \rightarrow -M_{ss}(\theta)$ and hence there exists the relation of symmetry
$K_{ij}(\theta) = \pm D_{ij}(\pi-\theta)$ with the $+$ or $-$ sign according to $i=j$ or $i \ne j$.
From the relation the measured quantities on the spin direction of the scattered or the recoil nucleon
in the laboratory system are found to have the symmetry relation $K_{ab}(\theta) = \pm D_{ab}(\pi-\theta)$ similarly.
Here, $a=n,S,L$ ($n$, $S$ and $L$ denote $\hat{y}$, $\hat{x}$ and $\hat{z}$ directions of the beam nucleon respectively) 
and $b=n,S,L$ (the $n$ denotes the $\hat{y}$ direction and the $S$ and $L$ denote $\hat{K}$ and $\hat{P}$ directions for  $D_{ab}(\theta)$ and $\hat{P}$ and $-\hat{K}$ directions for $K_{ab}(\theta)$ respectively) 
with the $+$ or $-$ sign according to $a=b$ or $a \ne b$.
Therefore breaking of the symmetry suggests occurrence of the unknown process 
out of the present framework of the formulation.
The same relation between $K_{ab}$ and $D_{ab}$ exists also in $p$-$p$ elastic scattering.
\\\hspace*{4.mm}
The spin transfer $K_{nn}(\theta) \equiv \frac{1}{2}\sum_{T=0,1} a_{0nn0}^T/I_0$ ($=D_t(\theta)$) at $\theta \ge 90^\circ$
is connected with $D_{nn}(\theta)$ at $\theta < 90^\circ$ as mentioned above.
In Fig. 8 results of the calculation are shown where the four curves are divided into two groups.
It is unexpected that $ps$ appears to describe the experimental data$\cite{McNaughton}$ better than $av$, however,
we have found the situation changes supposing the effect of the resonance.
While the spin singlet ($S$=$0$) $S$-wave is not influenced much in the $p$-$n$ elastic scattering,
for $P$-wave the resonance takes effect by modifying the phase shift parameter $\delta_L$ ($L$=1)  
as $\delta_1 \rightarrow \delta_1 \pm \pi/2$ 
so that the spin singlet element of the S matrix $S_1 = \rm{exp}(2 \it{i} \delta_{\rm 1}) $ reverses the sign.
With inclusion of it the theoretical curves $av$ and $ps$ seem to convert into another one implying 
the large discrepancies in the potential of these two equations. 
For ${}^1 P_1$ state $T$=$0$ is assigned and the energy is $\sim 500$ MeV the resonance may be caused 
by the two pion correlation in other words the $\sigma$ meson.
At the intermediate energy region it is supposed that the resonant process 
begins elastically via the quasi-stationary state $p+n \rightarrow \sigma+p+n$ in the spin singlet $P$-wave
as well as the usual one $\sigma$ meson exchange.
The procedure makes us understand the trend in general except for the observed oscillatory behavior of $K_{nn}(\theta)$.
\\\hspace*{4.mm}
The spin transfer $K_{LS}(\theta) \equiv \frac{1}{2}\sum_{T=0,1} a_{0Pz0}^T/I_0$ ($=A_t(\theta)$) 
is the parameter associated with the side direction of spin of the recoil nucleon and then unlike $K_{nn}(\theta)$ 
the additional angular dependence makes the role of the individual element of the M matrix be obscure somewhat.
Results of the calculation are still useful to study the $S$=$0$ part 
and the suggested resonant model of the meson-exchange force (Fig. 9).
In the $ps$ case as verified numerically the influence of $M_{ss}^0(\theta)$ on $K_{LS}(\theta)$ is relatively low
and so inclusion of the resonance changes the curve only a little, meanwhile, in the $av$ case the effect is obvious
particularly at $\theta \le 60^\circ$ and $\theta \ge 120^\circ$ 
where the $P$-wave ($\sim P_1(\theta)$) component of $S$=$0$ is enhanced.
Comparing with the measured data$\cite{Axen}$ at $\theta\ge60^\circ$ it seems that the $av$ case with the resonance
overestimates the data apart from $\theta\ge160^\circ$.
\\\hspace*{4.mm}
Similar to $K_{LS}(\theta)$ using the $av$ equation for the $S$=$0$ component $M_{ss}(\theta)$ 
the resonance has an effect on $K_{LL}(\theta) \equiv \frac{1}{2}\sum_{T=0,1} a_{0-Kz0}^T/I_0$ ($=-A^\prime_t(\theta)$) 
larger than the case of the $ps$ equation as shown in Fig. 10.
It is a problem that the effect results in an adverse effect particularly at $\theta=180^\circ$ in comparison with
the experimental data$\cite{McNaughton2}$.
The other way feasible to lower the value of $K_{LL}(\theta)$ at $\theta=180^\circ$ is to take the higher-order correction into account also for the ${}^3P_2$ state.
The calculation has already been done for $p$-$p$ elastic scattering$\cite{Kinpara}$, in which
the positive pion ($\pi^{+}$) is allowed to make the phase-shift a real number 
by substituting the $S$=$0$ $p$-$n$ scattering state for the $S$=$1$ $p$-$p$ state.
When we apply the procedure to $T_z$=$0$ case at the intermediate energy ($\sim 500$ ${\rm MeV}$) region,  
instead of pion the neutral component of the isovector vector meson ($\rho^0$) is supposed to take part in the process.
For the $av$ case without the resonance effect of ${}^1 P_1$, the correction of the ${}^3P_2$ state
reduces the values of $K_{LL}(\theta)$ roughly $\sim 0.1$ from the long-dashed curve 
toward the direction of the measured data at $\theta \ge 60^\circ$.
\\\hspace*{4.mm}
For spin transfer $K_{SS}(\theta) \equiv \frac{1}{2}\sum_{T=0,1} a_{0Px0}^T/I_0$ ($=R_t(\theta)$)
the effect of the ${}^3P_2$ higher-order correction becomes obvious since at $\theta\ge130^\circ$ the four curves of the calculation converge as shown in Fig. 11.
The cause of the trend is 
that the dependence of $I_0\,K_{SS}(\theta)$ on $M_{ss}^0(\theta)$ is represented by
${\rm Re}\,[M_{ss}^0(\theta)\,M_{1-1}^0(\theta)^\ast] \sim {\rm sin}\, 2\theta \, {\rm sin}\, \theta$ 
and so the ${}^1 P_1$ resonance does not work efficiently at the region.
In order to improve the discrepancy at $\theta\ge120^\circ$ we have added the correction of the ${}^3P_2$ state
for the $av$ case and have found it reduces the values also $\sim0.1$ at the region of $\theta$.
Furthermore, by the correction in conjunction with the ${}^1 P_1$ resonance the resulting curve turns to
decrease from the value $\sim 0$ at $\theta$\,=\,$120^\circ$ to $\sim -0.17$ at $\theta$\,=\,$180^\circ$.
In brief the ${}^3P_2$ correction enhances the $T$=$1$ sector of the M matrix
and then break from the ${\rm sin}\,2\theta \, {\rm sin}\,\theta$ dominance takes place 
so as to move the curve nearer to the experimental data$\cite{Axen}$. 
\\\hspace*{4.mm}
Since the correction of the ${}^3 P_2$ state is appropriate to scattering angles far from $\theta \sim 90^\circ$
it is expected to be effective also against the overestimate of the differential cross section.
In fact when the laboratory energy is 500 MeV the values at $\theta=0^\circ$ and $180^\circ$ reduce to $\sim 40\,\%$
of that in the $av$ case with no ${}^1 P_1$ resonance effect ($14.8 \times 0.4 \sim 6$) and which is empirically acceptable.
On the other hand when the energy is 200 MeV it changes to $\sim4.5$ at $\theta=0^\circ$ and $180^\circ$ 
giving too large effect and so an adjustment may be required to suppress the $\rho^0$ meson process 
for the energy region.
\\\hspace*{4.mm}
In the formulation by the BS equation there are several components
which are divided into two by the composite spin states subject to the potential.
To choose one spin singlet wave within two components a number of spin observables are useful.
Similar to $p$-$p$ elastic scattering the calculations result in the dominance of the axial-vector component
but the present situation may be changed by the investigation of the simultaneous equations in detail.
The resonant properties in the spin singlet state and the higher-order corrections for the spin triplet state
fill the remaining gap between the calculations and the experiments to some extent 
accompanying the virtual creation of mesons.

\newpage
\begin{figure}
\begin{center}
\scalebox{0.5}{\includegraphics{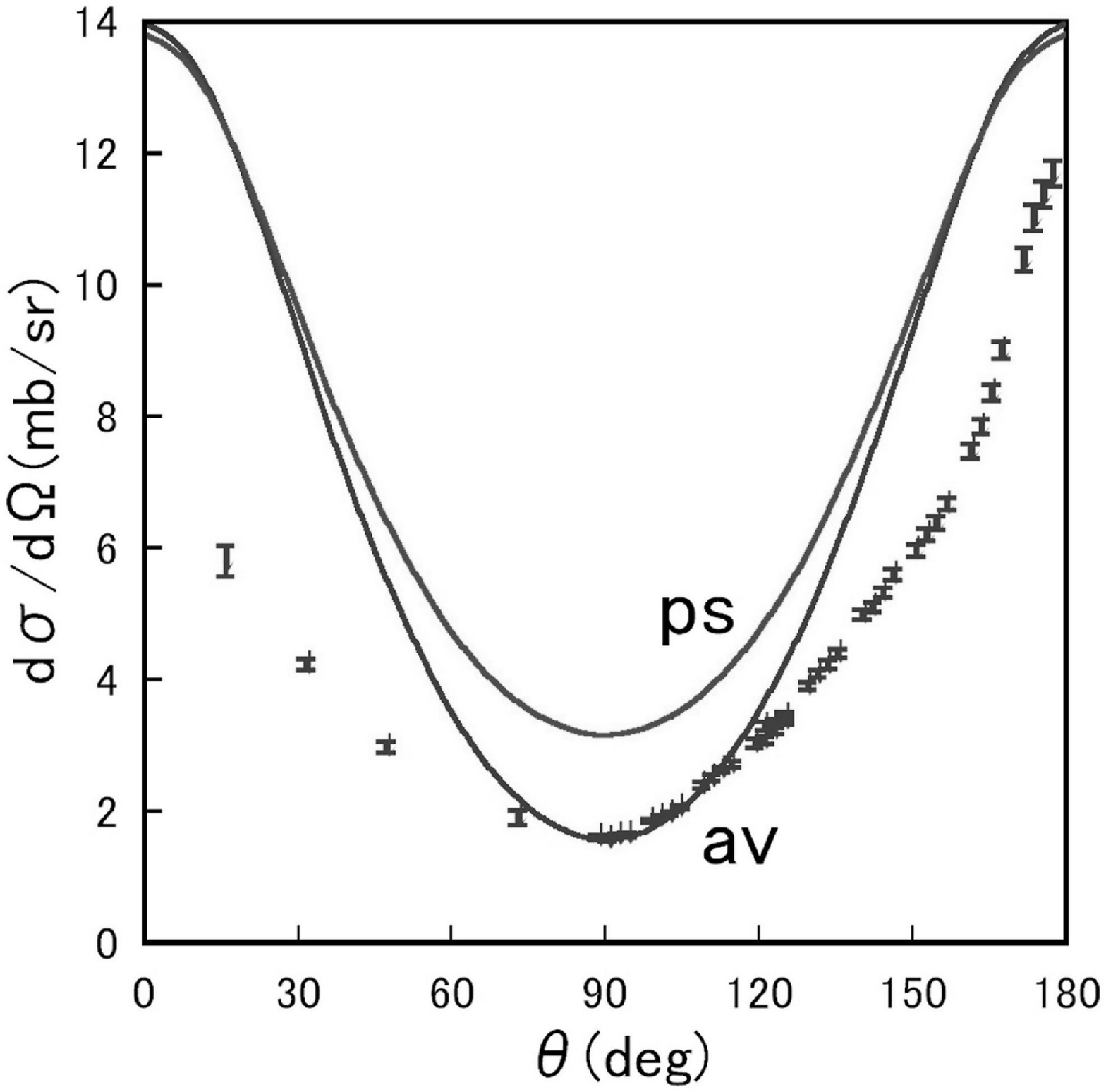}}
\caption{
The differential cross section $d \sigma /d \Omega$ (=$\,I_0$) as a function of the center of mass scattering angle $\theta$ at the laboratory energy of 200 MeV.
The $av$ and $ps$ denote the axial-vector and the pseudoscalar components for spin singlet part respectively.
The experimental data is from the 212 MeV in ref. $\cite{Keeler}$.
}
\end{center}
\end{figure}
\newpage
\begin{figure}
\begin{center}
\scalebox{0.5}{\includegraphics{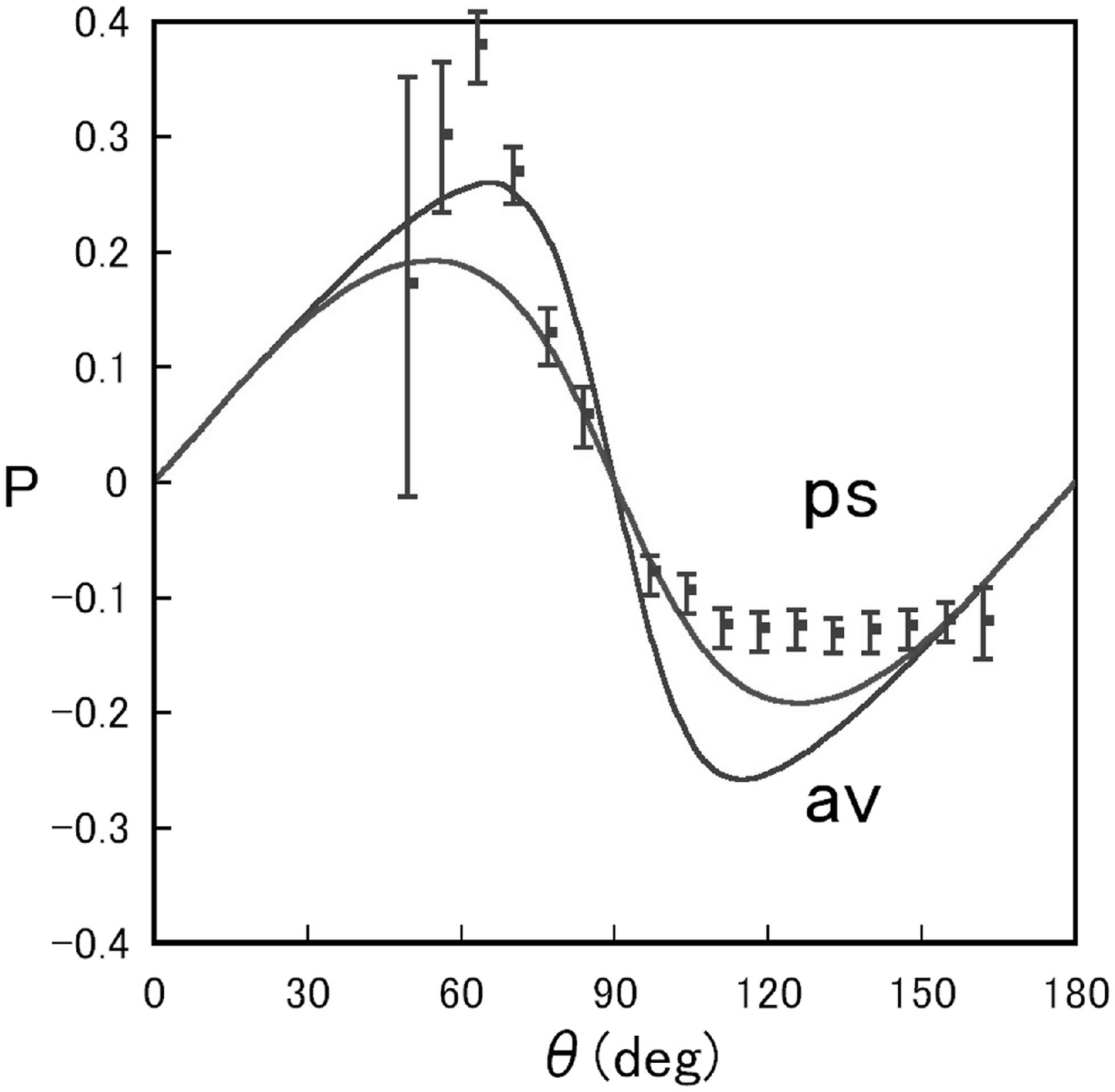}}
\caption{
The polarization $P$ (=$\,A_y$) as a function of the center of mass scattering angle $\theta$ at the laboratory energy of 200 MeV.
The $av$ and $ps$ denote the axial-vector and the pseudoscalar components for spin singlet part respectively.
The experimental data is from the 220 MeV in ref. $\cite{Clough}$.
}
\end{center}
\end{figure}
\newpage
\begin{figure}
\begin{center}
\scalebox{0.5}{\includegraphics{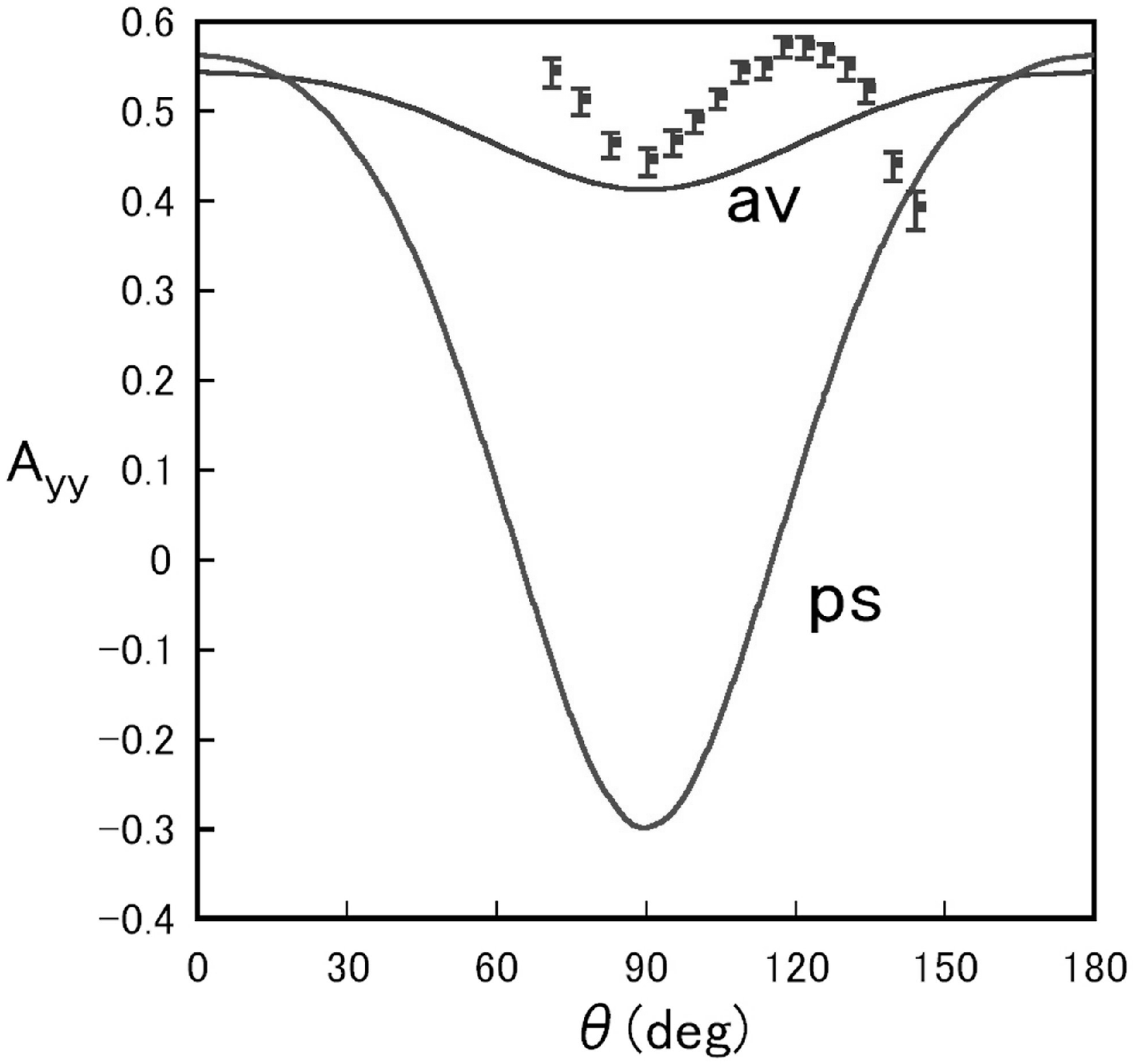}}
\caption{
The spin correlation parameter $A_{yy}$ as a function of the center of mass scattering angle $\theta$ at the laboratory energy of 200 MeV.
The $av$ and $ps$ denote the axial-vector and the pseudoscalar components for spin singlet part respectively.
The experimental data is from the 220 MeV in ref. $\cite{Bandyopadhyay}$.
}
\end{center}
\end{figure}
\newpage
\begin{figure}
\begin{center}
\scalebox{0.5}{\includegraphics{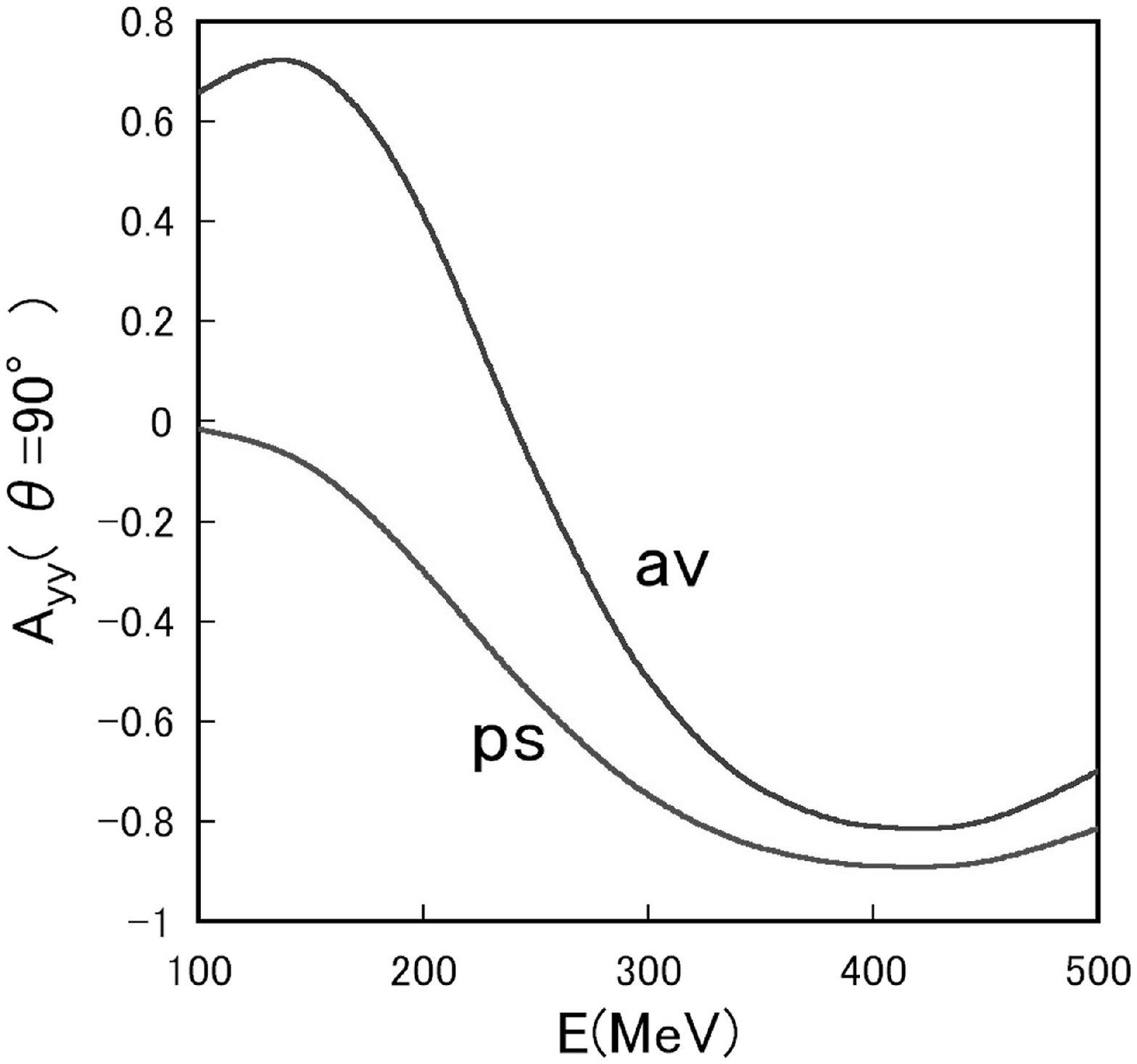}}
\caption{
The spin correlation parameter $A_{yy}$ at $\theta=90^\circ$ as a function of the laboratory energy $E$ in MeV.
The $av$ and $ps$ denote the axial-vector and the pseudoscalar components for spin singlet part respectively.
}
\end{center}
\end{figure}
\newpage
\begin{figure}
\begin{center}
\scalebox{0.5}{\includegraphics{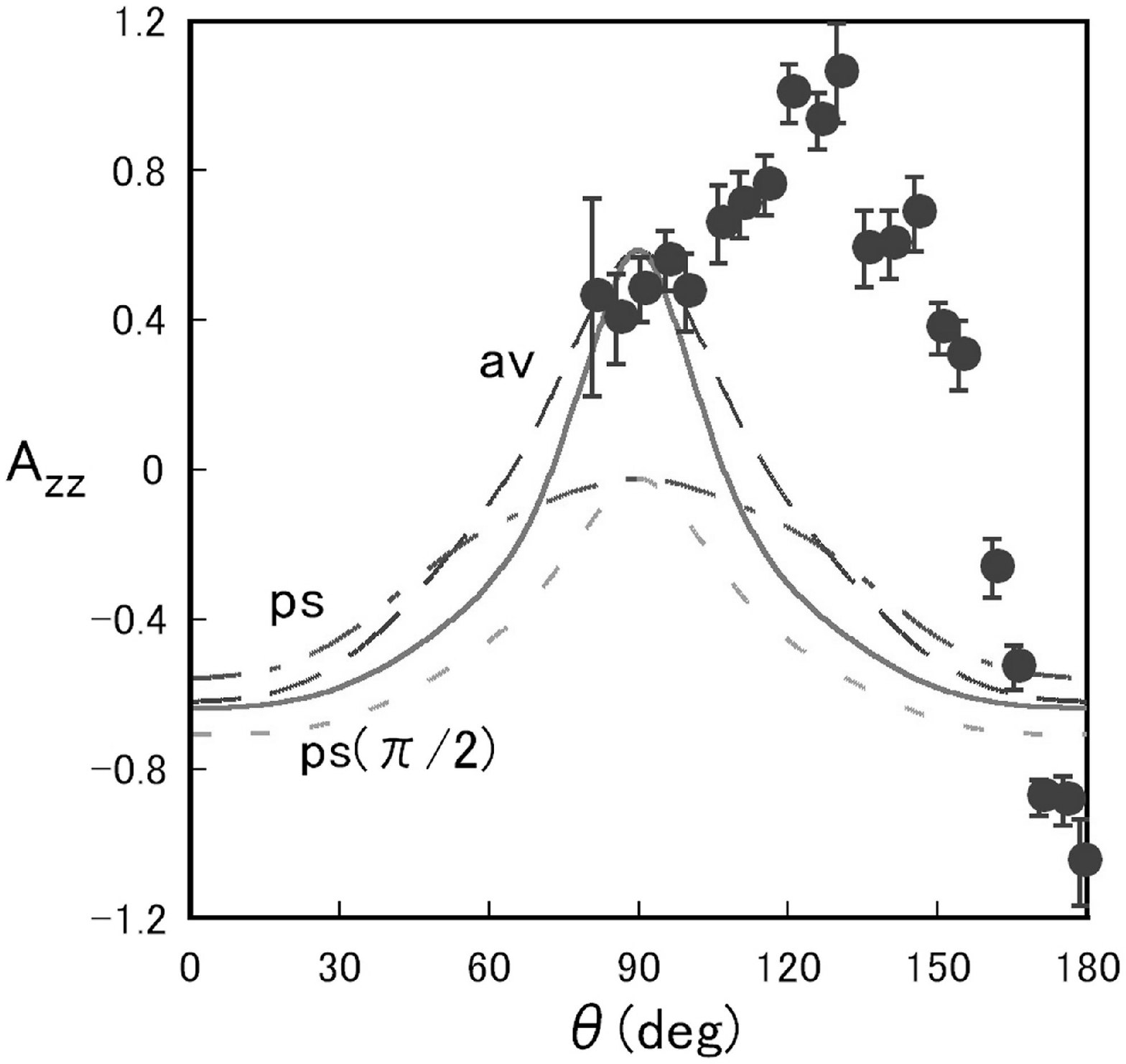}}
\caption{
The spin correlation parameter $A_{zz}$ as a function of the center of mass scattering angle $\theta$ at the laboratory energy of 500 MeV.
The $av$ (long-dashed) and $ps$ (dash-dot) denote the axial-vector and the pseudoscalar components 
for spin singlet part respectively.
The $av(\pi/2)$ (solid) and $ps(\pi/2)$ (short-dashed) denote the axial-vector and the pseudoscalar components 
for spin singlet part with the resonance effect respectively.
The experimental data is from the 484 MeV in ref. $\cite{Ditzler}$.
}
\end{center}
\end{figure}
\newpage
\begin{figure}
\begin{center}
\scalebox{0.5}{\includegraphics{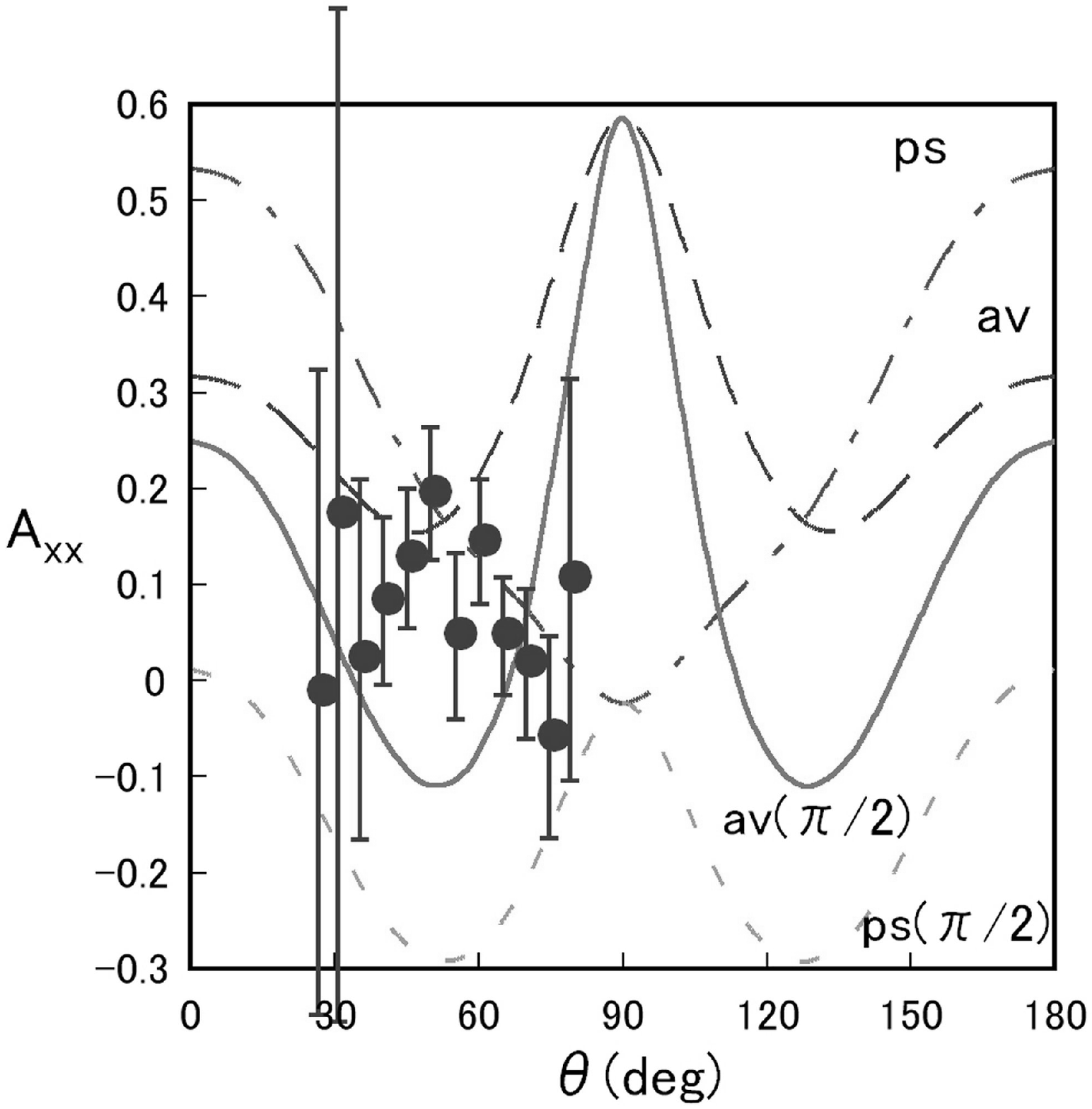}}
\caption{
The spin correlation parameter $A_{xx}$ as a function of the center of mass scattering angle $\theta$ at the laboratory energy of 500 MeV.
The $av$ and $ps$ denote the axial-vector and the pseudoscalar components for spin singlet part respectively.
The $av(\pi/2)$ and $ps(\pi/2)$ denote the axial-vector and the pseudoscalar components 
for spin singlet part with the resonance effect respectively.
The experimental data is from the 484 MeV in ref. $\cite{Shima}$.
}
\end{center}
\end{figure}
\newpage
\begin{figure}
\begin{center}
\scalebox{0.5}{\includegraphics{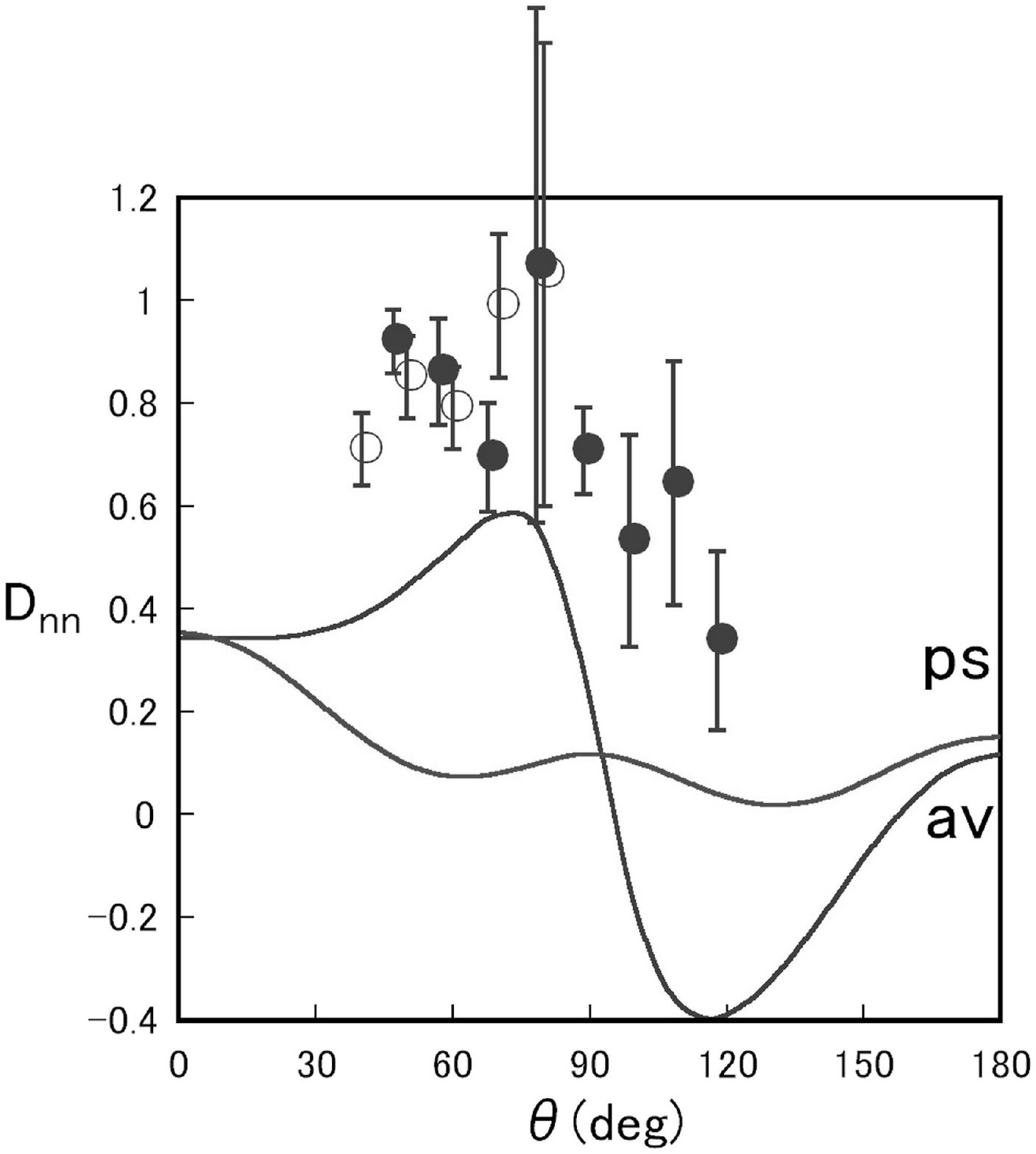}}
\caption{
The depolarization $D_{nn}$ as a function of the center of mass scattering angle $\theta$ at the laboratory energy of 300 MeV.
The $av$ and $ps$ denote the axial-vector and the pseudoscalar components for spin singlet part respectively.
The experimental data is from the 212 MeV in ref. $\cite{Warner}$ (open circle)
and the 647 MeV in ref. $\cite{Barlett}$ (solid circle).
}
\end{center}
\end{figure}
\newpage
\begin{figure}
\begin{center}
\scalebox{0.5}{\includegraphics{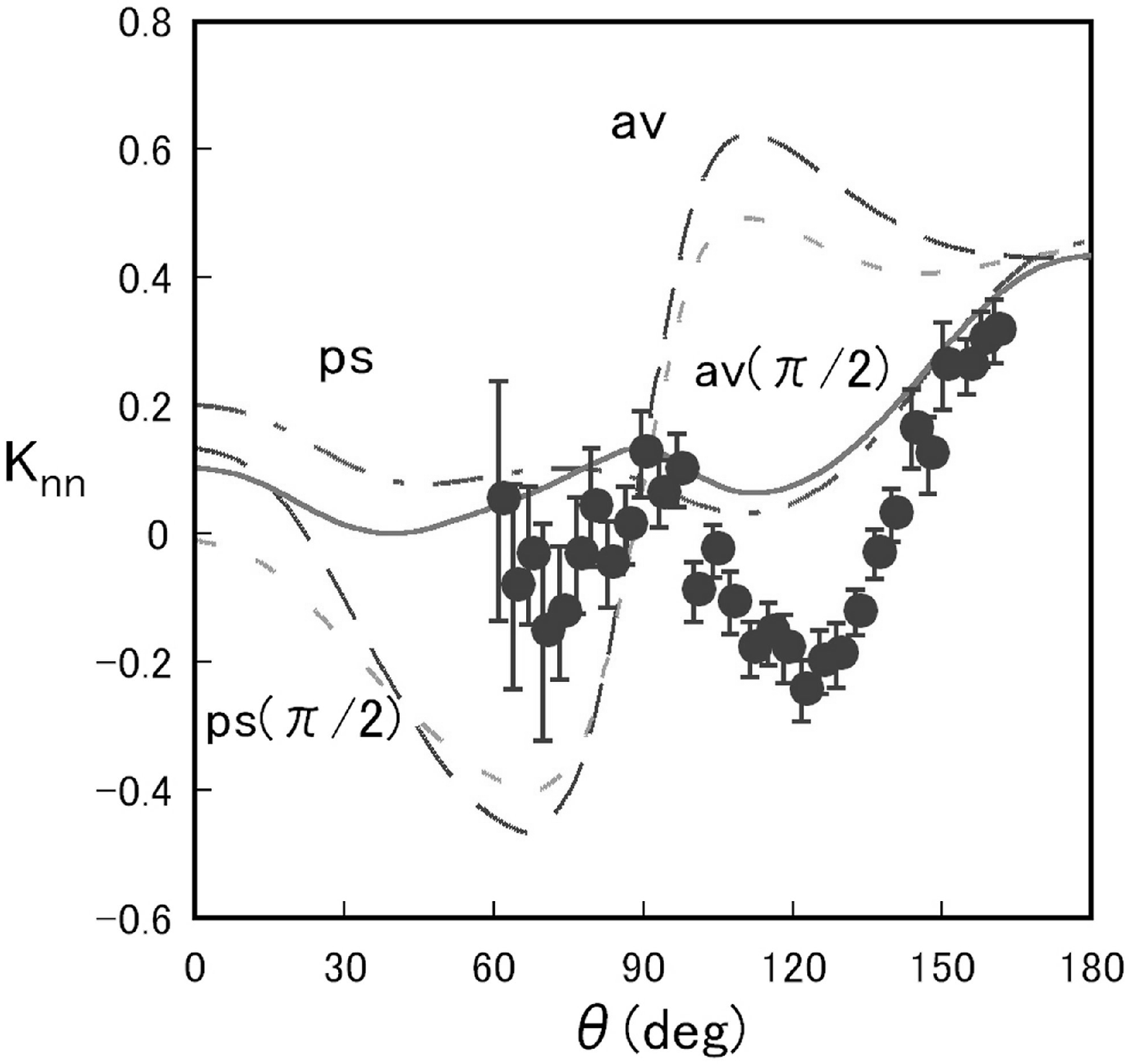}}
\caption{
The spin transfer $K_{nn}$ as a function of the center of mass scattering angle $\theta$ 
at the laboratory energy of 500 MeV.
The $av$ and $ps$ denote the axial-vector and the pseudoscalar components for spin singlet part respectively.
The $av(\pi/2)$ and $ps(\pi/2)$ denote the axial-vector and the pseudoscalar components 
for spin singlet part with the resonance effect respectively.
The experimental data is from the 485 MeV in ref. $\cite{McNaughton}$.
}
\end{center}
\end{figure}
\newpage
\begin{figure}
\begin{center}
\scalebox{0.5}{\includegraphics{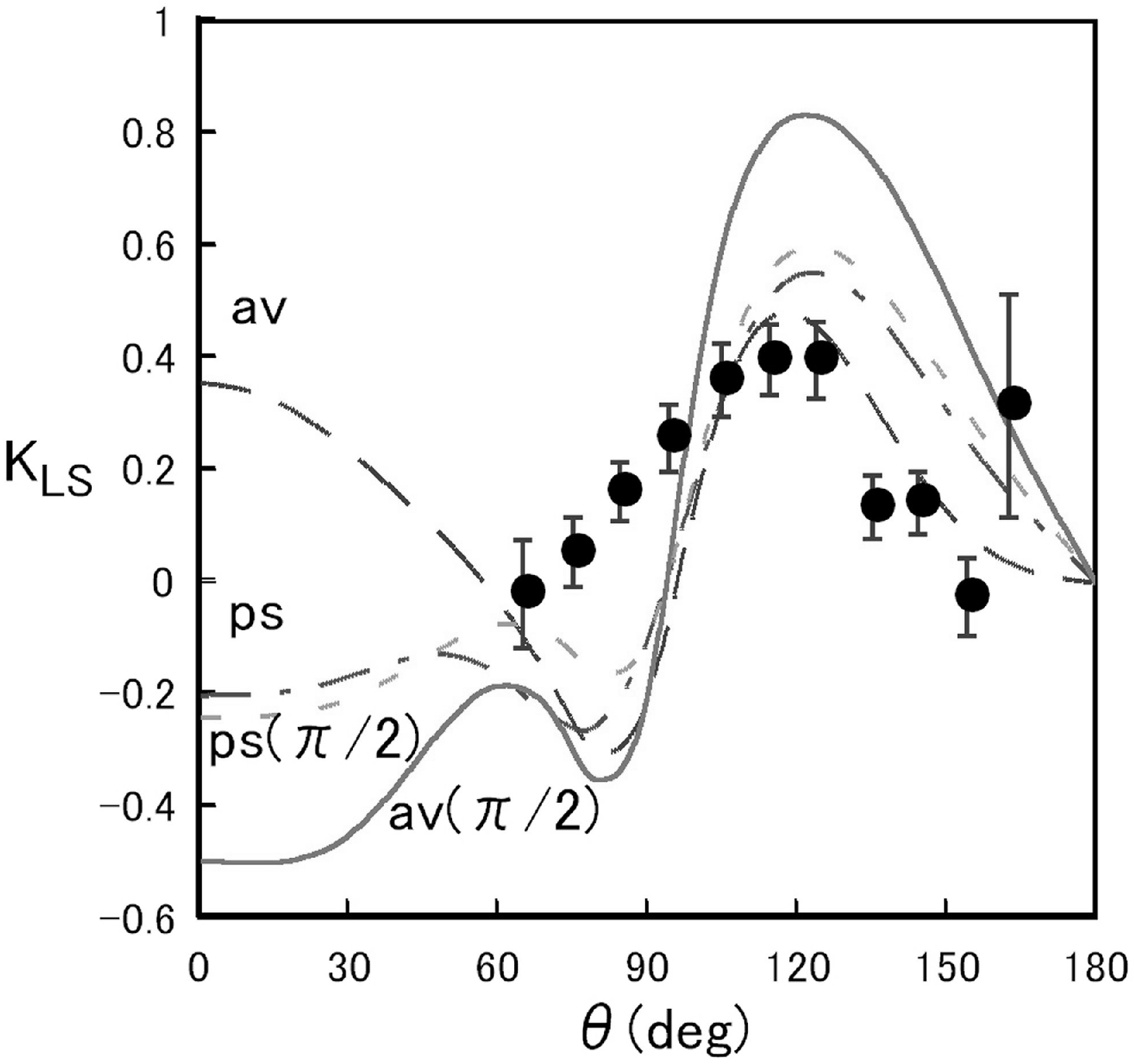}}
\caption{
The spin transfer $K_{LS}$ as a function of the center of mass scattering angle $\theta$ 
at the laboratory energy of 500 MeV.
The $av$ and $ps$ denote the axial-vector and the pseudoscalar components for spin singlet part respectively.
The $av(\pi/2)$ and $ps(\pi/2)$ denote the axial-vector and the pseudoscalar components 
for spin singlet part with the resonance effect respectively.
The experimental data is from the 495 MeV in ref. $\cite{Axen}$.
}
\end{center}
\end{figure}
\newpage
\begin{figure}
\begin{center}
\scalebox{0.5}{\includegraphics{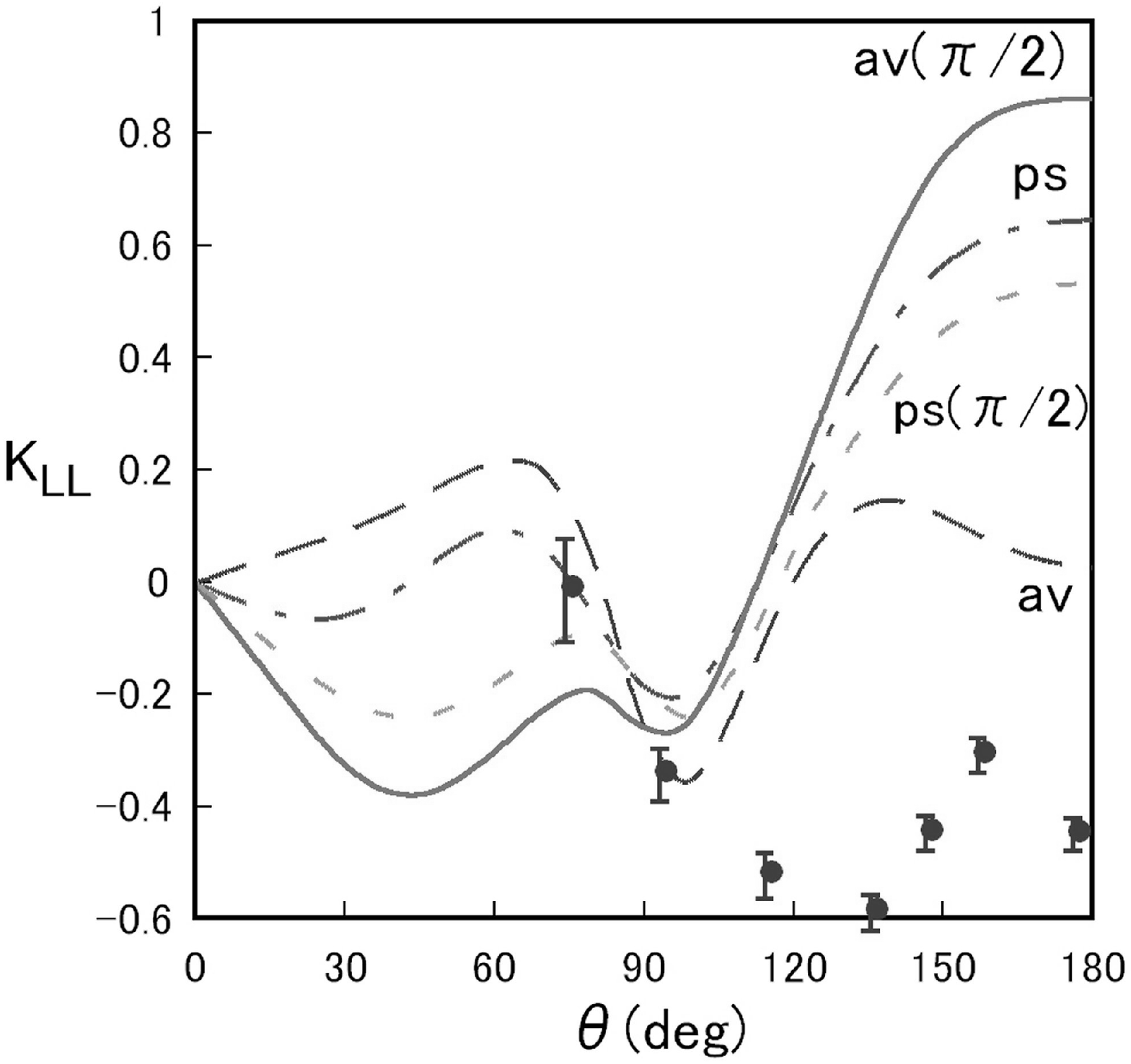}}
\caption{
The spin transfer $K_{LL}$ as a function of the center of mass scattering angle $\theta$ 
at the laboratory energy of 500 MeV.
The $av$ and $ps$ denote the axial-vector and the pseudoscalar components for spin singlet part respectively.
The $av(\pi/2)$ and $ps(\pi/2)$ denote the axial-vector and the pseudoscalar components 
for spin singlet part with the resonance effect respectively.
The experimental data is from the 485 MeV in ref. $\cite{McNaughton2}$.
}
\end{center}
\end{figure}
\newpage
\begin{figure}
\begin{center}
\scalebox{0.5}{\includegraphics{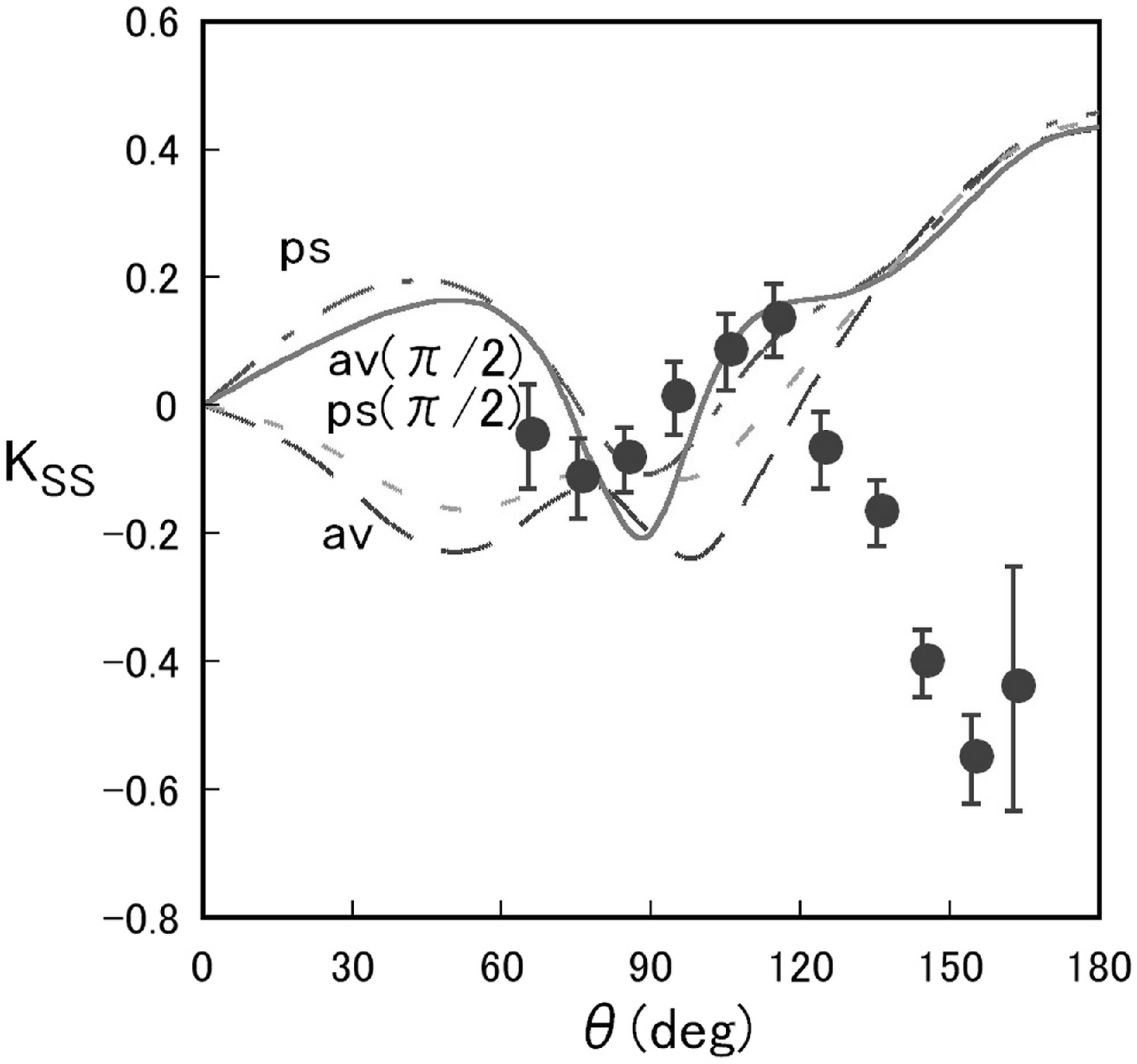}}
\caption{
The spin transfer $K_{SS}$ as a function of the center of mass scattering angle $\theta$ 
at the laboratory energy of 500 MeV.
The $av$ and $ps$ denote the axial-vector and the pseudoscalar components for spin singlet part respectively.
The $av(\pi/2)$ and $ps(\pi/2)$ denote the axial-vector and the pseudoscalar components 
for spin singlet part with the resonance effect respectively.
The experimental data is from the 495 MeV in ref. $\cite{Axen}$.
}
\end{center}
\end{figure}

\small
\begin{thebibliography}{99}
\bibitem{Kinpara}S. Kinpara, arXiv:nucl-th/1508.06393.
\bibitem{Keeler}R. K. Keeler, R. Dubois, E. G. Auld, D. A. Axen, M. Comyn, G. Ludgate, 
L. P. Robertson, J. R. Richardson, D. V. Bugg, J. A. Edgington, W. R. Gibson, A. S. Clough, N. M. Stewart and B. Dieterle, Nucl. Phys. {\bf A377}(1982)529.
\bibitem{Clough}A. S. Clough, D. R. Gibson, D. Axen, R. Dubois, L. Felawka, R. Keeler, G. A. Ludgate, C. J. Oram, C. Amsler,
D. V. Bugg, J. A. Edgington, L. P. Robertson, N. M. Stewart, J. Beveridge and J. R. Richardson,
Phys. Rev. {\bf C21}(1980)988.
\bibitem{Bandyopadhyay}D. Bandyopadhyay, R. Abegg, M. Ahmad, J. Birchall, K. Chantziantoniou, C. A. Davis, N. E. Davison,
P. P. J. Delheij, P. W. Green, L. G. Greeniaus, D. C. Healey, C. Lapointe, W. J. McDonald, C. A. Miller, G. A. Moss,
S. A. Page, W. D. Ramsay, N. L. Rodning, G. Roy, W. T. H. van Oers, G. D. Wait, J. W. Watson and Y. Ye,
Phys. Rev. {\bf C40}(1989)2684.
\bibitem{Ditzler}W. R. Ditzler, D. Hill, J. Hoftiezer, K. F. Johnson, D. Lopiano, T. Shima,
H. Shimizu, H. Spinka, R. Stanek, D. Underwood, R. G. Wagner, A. Yokosawa, G. R. Burleson,
J. A. Faucett, C. A. Fontenla, R. W. Garnett, C. Luchini, M. W. Rawool-Sullivan,
T. S. Bhatia, G. Glass, J. C. Hiebert, R. A. Kenefick, S. Nath, L. C. Northcliffe,
R. Damjanovich, J. J. Jarmer, J. Vaninetti, R. H. Jeppesen and G. E. Tripard,
Phys. Rev. {\bf D46}(1992)2792.
\bibitem{Shima}T. Shima, D. Hill, K. F. Johnson, H. Shimizu, H. Spinka, R. Stanek, D. Underwood, A. Yokosawa, G. Glass,
J. C. Hiebert, R. A. Kenefick, S. Nath, L. C. Northcliffe, G. R. Burleson, R. W. Garnett, J. A. Faucett, 
M. W. Rawool-Sullivan, R. Damjanovich, J. J. Jarmer, R. H. Jeppesen and G. E. Tripard,
Phys. Rev. {\bf D47}(1993)29.
\bibitem{Warner}R. E. Warner and J. H. Tinlot, Phys. Rev. {\bf 125}(1962)1028.
\bibitem{Barlett}M. L. Barlett, G. W. Hoffmann, L. Ray, G. Pauletta, K. H. McNaughton, J. F. Amann, K. W. Jones,
J. B. McClelland, M. W. McNaughton, R. Fergerson and D. Lopiano,
Phys. Rev. {\bf C40}(1989)2697.
\bibitem{McNaughton}M. W. McNaughton, K. Johnston, D. R. Swenson, D. Tupa, R. L. York, D. A. Ambrose, P. Coffey, 
K. H. McNaughton, P. J. Riley, G. Glass, J. C. Hiebert, R. H. Jeppesen, H. Spinka, I. Supek, G. E. Tripard and H. Woolverton,
Phys. Rev. {\bf C48}(1993)256.
\bibitem{Axen}D. Axen, R. Dubois, R. Keeler, G. A. Ludgate, C. J. Oram, L. P. Robertson, N. M. Stewart, 
C. Amsler, D. V. Bugg, J. A. Edgington, W. R. Gibson, N. Wright and A. S. Clough,
Phys. Rev. {\bf C21}(1980)998.
\bibitem{McNaughton2}K. H. McNaughton, D. A. Ambrose, P. Coffey, K. Johnston, P. J. Riley, M. W. McNaughton, K. Koch,
I. Supek, N. Tanaka, G. Glass, J. C. Hiebert, L. C. Northcliffe, A. J. Simon, D. J. Mercer, D. L. Adams, H. Spinka,
R. H. Jeppesen, G. E. Tripard and H. Woolverton,
Phys. Rev. {\bf C46}(1992)47.
\end{thebibliography}
\end{document}